\documentclass{article}
\usepackage{spconf,amsmath,graphicx}
\usepackage{booktabs}
\usepackage{multirow}
\usepackage{amssymb}
\usepackage{pifont}
\usepackage{hyperref}
\usepackage{color}
\usepackage[table]{xcolor}
\usepackage[numbers,sort]{natbib}
\definecolor{mygray}{RGB}{200,200,200}

\title{ICMC-ASR: The ICASSP 2024 In-Car Multi-Channel Automatic Speech Recognition Challenge}
\name{
    \begin{tabular}{c}
    \it He Wang$^1$, Pengcheng Guo$^1$, Yue Li$^1$, Ao Zhang$^1$, Jiayao Sun$^1$, Lei Xie$^{1{\ast}}$\thanks{*Corresponding author}, Wei Chen$^2$, Pan Zhou$^2$ \\
    \it Hui Bu$^3$, Xin Xu$^3$, Binbin Zhang$^4$, Zhuo Chen$^5$, Jian Wu$^6$, Longbiao Wang$^7$, Eng Siong Chng$^8$, Sun Li$^9$ \vspace{-0.25em}
    \end{tabular}
}
\address{$^1$Northwestern Polytechnical University~~$^2$Space AI, Li Auto~~$^3$Beijing AI Shell Technology Co. Ltd\\$^4$WeNet Open Source Community~~$^5$ByteDance~~$^6$Microsoft Corporation~~$^7$Tianjin University\\
$^8$Nanyang Technological University~$^9$China Academy of Information and Communication Technology\vspace{-0.25em}}

\makeatletter
\newcommand*{\rom}[1]{\expandafter\@slowromancap\romannumeral #1@}
\makeatother

\begin{document}
\ninept
\maketitle
\begin{abstract}
\vspace{-0.3em}
To promote speech processing and recognition research in driving scenarios, we build on the success of the Intelligent Cockpit Speech Recognition Challenge (ICSRC) held at ISCSLP 2022 and launch the ICASSP 2024 \textbf{I}n-\textbf{C}ar \textbf{M}ulti-\textbf{C}hannel \textbf{A}utomatic \textbf{S}peech \textbf{R}ecognition (ICMC-ASR) Challenge. 
This challenge collects over 100 hours of multi-channel speech data recorded inside a new energy vehicle and 40 hours of noise for data augmentation. 
Two tracks, including automatic speech recognition (ASR) and automatic speech diarization and recognition (ASDR) are set up, using character error rate (CER) and concatenated minimum permutation character error rate (cpCER) as evaluation metrics, respectively. 
Overall, the ICMC-ASR Challenge attracts 98 participating teams and receives 53 valid results in both tracks. 
In the end, first-place team USTC-iflytek achieves a CER of 13.16\% in the ASR track and a cpCER of 21.48\% in the ASDR track, showing an absolute improvement of 13.08\% and 51.4\% compared to our challenge baseline, respectively.

\end{abstract}
\vspace{-0.6em}
\section{Introduction}
\vspace{-0.5em}
\label{sec:intro}
Unlike common scenarios for automatic speech recognition (ASR) systems, such as home or meeting, the acoustic environment of the closed and irregular cockpit is complex.
In addition, various sources of noise exist during driving, such as wind, engine, wheel, car stereo, interfering speakers, etc. 
Therefore, how to leverage recent advances in speech processing and recognition to improve the robustness of in-car ASR systems is an essential problem worth investigating.

In 2022, we held the Intelligent Cockpit Speech Recognition Challenge~\cite{zhang2022iscslp} (ICSRC) and released a 20-hour single-channel evaluation set, collected in a hybrid electric vehicle,
While this dataset offers a valuable public testbed, there is still a lack of sizable real-world data to benchmark in-car ASR systems. 
To fill the gap, we build on the success of the previous ICSRC and launch the ICASSP 2024 In-Car Multi-Channel Automatic Speech Recognition (ICMC-ASR) Challenge, dedicated to the domain of speech processing and recognition in complex driving conditions. 
Furthermore, the ICMC-ASR dataset comprises an over 100h collection of real-world recorded, multi-channel, multi-speaker, in-car conversational Mandarin speech data and 40 hours of in-car recorded multi-channel noise audio.
We provide ASR and automatic speech diarization and recognition (ASDR) tracks targeting in-car multi-speaker chatting scenarios, using character error rate (CER) and concatenated minimum permutation character error rate (cpCER) as evaluation metrics, respectively. 
Finally, the ICMC-ASR Challenge attracts 98 teams to participate and receives 53 valid results in both tracks.
The USTC-iflytek team secures the championship in ASR Track with a CER of 13.16\% and ASDR Track with a cpCER of 21.48\%, showing remarkable improvements over the challenge baseline.

\begin{table}[t]
    \centering
    \caption{Statics about the ICMC-ASR dataset, including the duration of segmented near-field audio (Duration), number of sessions (Session), whether providing ground-truth speaker diarization (GT SD), transcriptions (Transcription) and near-field audio (Near-field).}
    \resizebox{\linewidth}{!}{
        \begin{tabular}{cccccc}
            \toprule
             Dataset  & Duration (h) & Session & GT SD & Transcription & Near-field \\
            \hline
            Train & 94.75 & 568 & \text{\ding{51}} & \text{\ding{51}} & \text{\ding{51}} \\
            Dev & 3.33 & 18 & \text{\ding{51}} & \text{\ding{51}} & \text{\ding{55}} \\
            Eval$_1$ & 3.30 & 18  & \text{\ding{51}} & \text{\ding{55}} & \text{\ding{55}} \\
            Eval$_2$ & 3.58 & 18 & \text{\ding{55}} & \text{\ding{55}} & \text{\ding{55}} \\
            Noise & 40.29 & 60 & - & - & - \\
            \bottomrule
        \end{tabular}
    }
    \label{dataset}
\end{table}

\begin{table*}[ht]
\centering
	\caption{Major techniques and results from top performing teams and our baseline. It is noteworthy that speech frontend and ASR backbone are applied in both tracks, while speaker diarization is \textbf{only} for Track \rom{2}. Bolded teams are invited to submit a paper to ICASSP 2024.}
    \label{table-results}
    \resizebox{\linewidth}{!}{
        \begin{tabular}{c|ccc|cc}
            \hline
            \multirow{2}{*}{Team} & \multicolumn{3}{c|}{Track \rom{1}} & \multicolumn{2}{c}{Track \rom{2}}\\
            \cline{2-4} \cline{5-6} ~ & Speech Frontend & ASR Backbone & CER (\%) $\downarrow$ & Speaker Diarization & cpCER (\%) $\downarrow$\\
            \hline
            \textbf{USTC-iflytek} & Modified GSS~\cite{wang2023ustcnercslip} & Accent-ASR & \cellcolor{mygray} 13.16 (1st) & MC-TS-VAD~\cite{wang2023ustcnercslip} & \cellcolor{mygray} 21.48 (1st) \\
            \textbf{Fosafer Research} & AEC + IVA, Beamforming, MP-SENet~\cite{lu2023mpsenet} &  HuBERT~\cite{hsu2021hubert} + Conformer & \cellcolor{mygray} 14.63 (2nd) & SincNet and LSTM based VAD & \cellcolor{mygray} 29.33 (5th) \\
            \textbf{FawAISpeech} & WPE, GSS  & MC-E-Branchformer & \cellcolor{mygray} 14.72 (3rd) & - & \cellcolor{mygray} - \\ 
            Nanjing Longyuan & AEC + IVA, DEEP-FSMN~\cite{zhang2018deep} &  Data2vec2~\cite{baevski2023efficient}+ E-Branchformer & \cellcolor{mygray} 15.62 (4th) & ECAPA-TDNN-1024~\cite{desplanques2020ecapa} VAD & \cellcolor{mygray} 26.48 (4th) \\ 
            \textbf{RoyalFlush} & AEC + IVA, WPE, GSS, DCCRN-VAE~\cite{xiang2023deep} & HuBERT + E-Branchformer & \cellcolor{mygray} - & Resnet34-TSDP~\cite{tian2022royalflush} based TS-VAD & \cellcolor{mygray} 25.88 (2nd) \\ 
            \textbf{Ximalaya Speech} & AEC + IVA & HuBERT + E-Conformer & \cellcolor{mygray} - & Resnet34 based TS-VAD & \cellcolor{mygray} 26.37 (3rd) \\ 
            HLT2023-NUS & AEC + IVA, WPE, GSS & HuBERT + E-Branchformer & \cellcolor{mygray} - & CAM++~\cite{wang2023cam++} based TS-VAD & \cellcolor{mygray} 31.68 (6th) \\ 
            Baseline & AEC + IVA & E-Branchformer & \cellcolor{mygray} 26.24 & Pyannote VAD\footnotemark \ & \cellcolor{mygray} 72.88 \\ 
            \hline
        \end{tabular}
    }
    \vspace{-2.0em}
\end{table*}

\vspace{-1em}
\section{Challenge Description}
\vspace{-0.7em}
\subsection{Dataset}
\vspace{-0.5em}
The dataset utilized by ICMC-ASR Challenge is collected in a hybrid electric vehicle with speakers sitting in different positions, including the driver seat, passenger seat, and two rear seats.
Specifically, there are 4 distributed microphones placed at four seats to record every speaker, constituting the far-field data. 
For transcription purposes, each speaker wears a high-fidelity headphone to record near-field data. 
As the realistic acoustic environment of driving scenarios is complex and involves a variety of noises, we carefully design the recording environments to ensure comprehensive coverage. 
Specifically, we achieve this by varying driving-related factors, including driving road (downtown street and highway), vehicle speed (parking, slow, medium and fast), air-conditioner (off, medium and high), music player (off and on), driver-side window and sunroof (close, open one-third and open halfway), driving time (daytime and nighttime).
By combining these factors, we finally form 60 different scenarios covering most in-car acoustic environments.

Overall, the ICMC-ASR dataset contains over 100 hours (near-field audio by oracle timestamps) of in-car chatting data in total, which is divided into training (Train) set, development (Dev) set, evaluation sets for ASR Track (Eval$_1$) and ASDR Track (Eval$_2$). 
For each set, far-field audio of 4 channels is included, but only the Train set will contain near-field audio. 
Particularly, for the Eval$_1$ set, oracle timestamps will be available, whereas for Eval$_2$, it is not available, requiring participants to utilize speaker diarization techniques for audio segmentation. 
Additionally, a sizable noise dataset (Noise) is provided, following the recording setup of the far-filed data but without speaker-talking.
The detailed information for each data subset is available in Table \ref{dataset}.

\vspace{-1em}
\subsection{Track setting and evaluation}
\textbf{Track \rom{1} Automatic Speech Recognition (ASR)}: 
In this track, participants are provided with the oracle segmentation of the evaluation set. 
The primary objective of this track is to focus on the development of ASR systems based on multi-channel multi-speaker speech data. 
Participants need to devise algorithms that can effectively fuse information across different channels, suppress inevitable noise, handle multi-speaker overlaps, etc.
For this track, the accuracy of the ASR system is measured by character error rate (CER). 

\textbf{Track \rom{2} Automatic Speech Diarization and Recognition (ASDR)}: 
Unlike Track \rom{1}, Track \rom{2} does not provide any prior or oracle information during evaluation (e.g. segmentation and speaker label for each utterance, total number of speakers in each session, etc.). 
Participants in this task are required to design automatic systems for both speaker diarization and transcription. 
For this track, we adopt concatenated minimum permutation character error rate (cpCER) as the metric for ASDR systems. 

\section{Results and Discussion}
\addtocounter{footnote}{0}
\footnotetext{\url{https://github.com/pyannote/pyannote-audio}}
Table \ref{table-results} shows major techniques and results from top-performing teams of this challenge, as well as our baseline\footnote{\url{https://github.com/MrSupW/ICMC-ASR_Baseline}}, implemented on the WeNet toolkit~\cite{yao2021wenet}. 
For a complete leaderboard and detailed system reports, please refer to our official website\footnote{\url{https://icmcasr.org}}. 
We calculated the CER and cpCER metrics for submissions from 35 teams of Track \rom{1} and 18 teams of Track \rom{2}, respectively. 
According to these, the champions of both tracks in the ICASSP 2024 ICMC-ASR Challenge are the team USTC-iflytek, achieving outstanding scores of CER 13.16\% and cpCER 21.48\% on Track \rom{1} and Track \rom{2}. 
Next, we will proceed with the discussion of effective techniques, delving into three aspects: speech frontend,  ASR backbone and speaker diarization.

\textbf{Speech Frontend.} 
Most teams use Acoustic Echo Cancellation (AEC) and Independent Vector Analysis (IVA) following our challenge baseline. 
Some teams additionally incorporate weighted prediction error (WPE) and guided source separation (GSS), for example, teams RoyalFlush, FawAISpeech, and HLT2023-NUS. 
Moreover, neural network (NN) based speech frontend denoising models, including MP-SENet~\cite{lu2023mpsenet}, DCCRN-VAE~\cite{xiang2023deep}, and DEEP-FSMN~\cite{zhang2018deep}, are favored by many participating teams. 
Especially, team USTC-iflytek utilizes energy and phase differences instead of the traditional maximum signal-noise ratio (SNR) criterion in GSS for channel selection, while a recursive smoothing technique in the beamformer to assess power spectral density matrices, providing higher quality single-channel audio for the downstream ASR.

\textbf{ASR Backbone.}
According to the rule of no external text data, many teams chose to generate audio features using self-supervised learning (SSL) models, fed into mainstream ASR models for training. 
The HuBERT~\cite{hsu2021hubert} SSL model is the most popular one, used by 4 teams. 
While team Nanjing Longyuan uses Data2vec2~\cite{baevski2023efficient} and introduces online noise enhancement techniques. 
Differently, team USTC-iflytek iteratively generates pseudo-labels for unlabeled data and proposes an Accent-ASR model optimized for accents. 
Team FawAISpeech introduces a multi-channel ASR model based on E-Branchformer and cross attention~\cite{guo2023npu}, also without an SSL model.

\textbf{Speaker Diarization.} 
Most teams improve upon the TS-VAD~\cite{medennikov2020target} for the speaker diarization module, except teams Fosafer Research and Nanjing Longyuan, which use NN-based VAD models. 
Team USTC-iflytek extends the TS-VAD model to multi-channel audio, proposing and utilizing the Multi-Channel TS-VAD~\cite{wang2023ustcnercslip}. 
Teams RoyalFlush, Ximalaya Speech, and HLT2023-NUS follow a similar approach, using different NN-based models to extract speaker embedding to replace the i-vector used in traditional TS-VAD.

In the end, we would like to express our gratitude to all participating teams for their contributions during the ICMC-ASR challenge and believe that our efforts contribute to advancing human-vehicle interaction towards greater accuracy and convenience.

\bibliographystyle{IEEEbib}
\bibliography{refs}

\end{document}